# Twisted oxide membrane interface by local atomic registry design

Min-Su Kim[1†], Kyoungjun Lee[2†], Ryo Ishikawa[3†], Kyung Song[4], Naafis Ahnaf Shahed[5], Ki-Tae Eom[2], Mark S. Rzchowski[6], Evgeny Y. Tsymbal[5], Naoya Shibata[3], Teruyasu Mizoguchi[7*], Chang-Beom Eom[2*], and Si-Young Choi[1,8,9*]

[1]*Department of Materials Science and Engineering, Pohang University of Science and Technology (POSTECH), Pohang 37673, Republic of Korea*

[2]*Department of Materials Science and Engineering, University of Wisconsin-Madison, Madison, WI 53706, United States of America*

[3]*Institute of Engineering Innovation, The University of Tokyo, Tokyo 113-8656, Japan*

[4]*Department of Materials Analysis, Korea Institute of Materials Science (KIMS), Changwon 51508, Republic of Korea*

[5]*Department of Physics and Astronomy & Nebraska Center for Materials and Nanoscience, University of Nebraska, Lincoln, NE 68588, United States of America*

[6]*Department of Physics, University of Wisconsin-Madison, Madison, WI 53706, United States of America*

[7]*Institute of Industrial Science, The University of Tokyo, Tokyo 153-8505, Japan*

[8]*Department of Semiconductor Engineering, Pohang University of Science and Technology (POSTECH), Pohang 37673, Republic of Korea*

[9]*Center for Van der Waals Quantum Solids, Institute for Basic Science (IBS), Pohang 37673, Republic of Korea*

[†]*These authors contributed equally: Min-Su Kim (M.-S.K.), Kyoungjun Lee (K.L), Ryo Ishikawa (R.I.)*

*\*Corresponding author. Email: youngchoi@postech.ac.kr (S.-Y.C.); ceom@wisc.edu (C.B.E.); teru@iis.u-tokyo.ac.jp (T. M.);*

# Abstract

Interplay of lattice, orbital, and charge degrees of freedom in complex oxide materials has hosted a plethora of exotic quantum phases and physical properties[1-3]. Recent advances in synthesis of freestanding complex oxide membranes[4,5] and twisted heterostructures assembled from membranes[6-9] provide new opportunities for discovery using moiré design with local lattice control. To this end, we designed moiré crystals at the coincidence site lattice condition, providing commensurate structure within the moiré supercell arising from the multi-atom complex oxide unit cell. We fabricated such twisted bilayers from freestanding $SrTiO_3$ membranes and used depth sectioning-based TEM methods to discover ordered charge states at the moiré interface. By selectively imaging $SrTiO_3$ atomic planes at different depths through the bilayer, we clearly resolved the moiré periodic structure at the twisted interface and found that it exhibits lattice-dependent charge disproportionation in the local atomic registry within the moiré supercell. Our density-functional modelling of the twisted oxide interface predicts that these moiré phenomena are accompanied by the emergence of a two-dimensional flat band that can drive new electronic phases. Our work provides a novel guideline for controlling moiré periodicity in twisted oxides and opens pathways to exploit the new functionalities via moiré lattice-driven charge-orbital correlation.

# Introduction

Understanding the lattice, charge, and orbital interactions[1] in complex oxide materials has been essential in elucidating novel electronic phases such as superconductivity[2,10] and two-dimensional electron gases[3,11,12] at oxide interfaces. Such interfacial electronic reconstructions can arise from localized states crossing the Fermi level, for instance, an electronic potential induced electron pocket[13-16]. Such electronic potential wells can also arise from the periodically modulated electronic structure at moiré interfaces in twisted bilayers of 2D van der Waals materials[17], tuned by the twist angle. Complex oxides provide new directions for interfacial electronic phases. The different bonding strengths and directions in oxides compared to 2D materials lead to different hybridized orbital structures[18,19], and the complex unit cell provides new opportunities for moiré design. The resulting interfacial electronic and lattice structure brings new phenomena to twisted bilayers.

Freestanding membranes of epitaxial oxide thin films form the basis for assembling such twisted bilayers. Freestanding oxide membranes have been recently fabricated using sacrificial buffer layers[4,5,20-23] and brings transferability opportunities comparable to 2D van der Waals materials. Advances in transfer techniques have led to the potential for high-quality twisted oxide bilayer heterostructures[6-8], but guidelines for interface design and fundamental understanding based on complex oxides interacting with moiré lattices have not been previously developed. Here, we provide theory-based design strategies and investigate these with high-quality twisted membrane heterostructure synthesis and microscopic lattice and electronic structure investigations.

Strontium titanate, $SrTiO_3$ (STO), is an excellent candidate membrane system whose electronic structure and lattice properties have been well-studied. It is an insulating oxide with empty Ti $d$-orbitals crystal field split into $t_{2g}$ and $e_g$ states[24] – these orbital states can be engineered for interfacial lattice modulation and to host electrons from artificial charge transfer [12,25]. Even in the form of freestanding membrane, STO exhibits the enhanced ease of crystal structure modulation and thereby lattice-induced electronic tunability[26,27]. We use high-resolution scanning transmission electron microscopy (STEM) to directly observe the moiré superlattice in STO twisted bilayer membrane assembled heterostructures and discover moiré 2D spatial charge modulation with layer-resolved STEM electron energy loss spectroscopy (EELS). We show that this charge modulation arises from local structural differences at the interface which we classify within a coincident site lattice (CSL) framework. Our first-principles calculations explain the lattice-dependent charge disproportionation at the twisted oxide interface consistent with the experimental results and predict the emergence of interface confined flat bands. These results reveal that unprecedented moiré phenomena can occur at twisted oxide bilayers and provide opportunities for moiré engineering of novel electronic phases.

**Moiré interface design**

An ingredient to exploiting moiré superlattices in correlated oxides is the complex repeating unit cell[28,29]. The twist angle between the crystals not only determines the repeating moiré supercell at the interface but also within the moiré supercell at specific angles aligns atomic columns between membranes. This forms a commensurate lattice known as the coincidence site lattice (CSL)[30]. These are low energy interfaces, labelled by a $\Sigma$ value characterizing the degree of commensurability[31,32]. Different atomic coordination and bonding at unique CSL points in the moiré superlattice could suppress energy band dispersion and induce charge localization[33,34]. We explore such moiré electronic reconstruction through strong lattice-electron correlations.

We designed a SrTiO₃ (STO) membrane bilayer with a twist angle of 10.4° (Σ 61) and a 3.04 nm periodic moiré supercell (**Fig. 1a**). The oxide bilayer has the stoichiometric interfacial geometry terminated by SrO plane for top membrane and TiO₂ plane for bottom membrane, respectively, based on the ternary ABO₃ perovskite structure of STO. The three non-equivalent atoms in STO form three unique CSLs, where Sr, Ti, and O align in columns at different commensurate points within the moiré cell, in addition to an incommensurate point (non-CSL), as shown in **Fig. 1a**.

At Ti-CSL points, four Sr atoms surround the Ti ion in the upper and lower layers, replicating the Ti-O octahedral structure characteristic of bulk STO at the interface (**Fig. 1b**). Similarly, the Ti-O polyhedral structure around Sr-CSL sites form octahedral structures where the Ti ion coordinates with six oxygen atoms (**Fig. 1c**). Since both types of cation-centered CSLs exhibit minimal structural differences at the interface, there is little change in Ti oxidation states. However atomic rearrangements at some local CSL structures can exceed the limits for octahedral coordination, leading to different properties[35]. In our case, this leads to a lower coordination number (CN), changing the charge state of Ti. Considering the O-CSL point centered around a corner-sharing oxygen atom, it is evident that the oxygen atom at the interfacial Ti-O polyhedron remains 5-fold coordinated due to the absence of a structural connection with the SrO layer in the top STO (**Fig. 1d**). As a result, this O-CSL point exhibits a relatively low coordination number of CN = 5, and the Ti-O polyhedron in non-CSL structure with atomically staggered distributions also coordinates with only five oxygen atoms (**Fig. 1e**).

We design unique Ti-O coordination differences within each CSL structure and identify changes in electronic structure according to local atomic structure using the CSL framework. To visualize the local lattice-driven charge state along the CSL point at the interface, we calculated the Bader charge[36] of Ti ions in the moiré supercell which is expected to correlate with the Ti oxidation state[37] (**Fig. 1f**). The Ti-CSL and Sr-CSL regions exhibit Bader charge values consistent with that of bulk STO, having the similar coordination. However, the O-CSL and non-CSL regions exhibit smaller Bader charge values indicating a reduced Ti ion oxidation state due to the reduced coordination numbers. This charge disproportionation is intrinsic to the designed moiré cell and determined by the local atomic registry. As the result, through the twisting of oxide interface, we can manipulate the in-plane charge disproportionation by the CSL/moiré lattice engineering.

**Experimental investigation**

We assembled STO twisted bilayers from epitaxial STO membranes released from their substrates with sub-degree angle control. The 5 nm-thick STO membranes were grown on and released from the $TiO_2$-terminated STO (001) substrate with an SCAO release layer, as described in Methods. The crystal quality and atomic surface flatness of the membranes were identified by in-situ reflection high-energy electron diffraction and topographic imaging (**Extended Data Fig. 1**). For the twisted bilayer STO membrane system, we repeated the SCAO etching and transfer process. We adopted a stage-rotational controller with 0.1° of step to precisely control the target twist angle and build the 10.4° CSL boundary (**Fig. 2a**). The fabrication process is described in **Extended Data Fig. 2**. We obtained an optically wrinkle/crack free large area of the twisted membrane, as shown in **Fig. 2b**. And we can clearly see the atomically flat step-terrace structure of both top and bottom membrane, as shown in the topographic image of **Fig. 2c**. The height difference between the top and bottom membrane was ~ 5 nm corresponding to the film thickness, as we observed by cross-sectional STEM, and the step height was ~ 4 Å corresponding to the unit cell height of STO (**Fig. 2d**). Furthermore, the moiré pattern and twist angle in the twisted STO bilayer transferred on the TEM grid are confirmed by high-resolution (HR) TEM and electron diffraction pattern (**Extended Data Fig. 3a, b**). With precise angle control, we obtain the TEM sample with 10.4° of twist angle and 3.04 nm of moiré supercell size consistent with the theoretical values (**Extended Data Fig. 3c**). The substrate $TiO_2$ termination led to a membrane SrO termination on the substrate side and $TiO_2$ termination on the opposing side. The resulting interface had the SrO-$TiO_2$ atomic geometry, as we designed in **Fig. 1** (**Fig. 2e**). We fabricated various bilayers with different twist angles and membrane thicknesses.

We focus here on twist angles near 10°. This 10.4° twist angle targets the coincidence site lattice (CSL) condition $\Sigma$ 61 arising from this rotation between layers. We first conducted cross-sectional atomic-resolution STEM observations along three different projections as shown in **Fig. 2e** to determine interfacial bonding between the twisted membranes. The top aligned image reveals a 5 nm thick top STO membrane with an atomically flat termination along the [001] zone axis (**Fig. 2f**). The bottom twisted STO membrane is not aligned along the zone axis and shows linear contrast due to overlapping atomic structure. The interface structure has a measured periodicity of 4.42 nm consistent with a 10.4° twist angle moiré periodicity of 3.04 nm and 45° rotation between the [100] zone axis and moiré cell. The ~0.39 nm lattice parameters near the interface are in good agreement with the standard unit cell length of cubic STO. The magnified HAADF image clearly shows the atomic contrast, and its line profile robustly indicates electron density at the SrO layer position in the moiré interface. This shows that the freestanding STO membranes interact with each other via atomic bonding.

Tilting the twisted membrane bilayer by ~5° around the [001] axis from the top STO membrane leads to linear contrast observed across all regions, including the interface, as all atomic layers appear twisted when compared with the top aligned axis (**Fig. 2g**). The bottom aligned region tilting with 10.4° from the zone axis also shows interface bonding with a lattice constant of ~ 4 Å, similar to that of the bulk (**Fig. 2h**). The atomically bonded twisted STO membrane preserved its atomic flatness when observed along various axes after the transfer process.

These cross-sectional STEM measurements quantify properties projected along the moiré interface. To determine the 2D bonding and charge density across the plane of the interface, we performed focus-controlled HAADF-STEM with depth sectioning (Methods) in a plan-view geometry. We used large-angle illumination STEM depth sectioning (Methods): a focal series of HAADF-STEM images were acquired along the plan-view of the 5 nm/ 5 nm twisted STO. Since the depth resolution in STEM is usually not enough to directly observe the atomic structure of the buried interface, we performed a large-illumination angle of 50 mrad at 300 kV, significantly improving the depth resolution[38]. By controlling the focus step of 1.16 Å, we selectively imaged the atomic structures of the top STO layer, bottom STO layer, and their interface. **Fig. 3a** shows seven such layers arranged according to their depth (**Extended Data Fig. 4** shows the same images without the 3D perspective).

From **Fig. 3a**, we can clearly see the atomic structures at the respective foci and directly confirm that the structures at 0 and 10 nm are twisted ~10° in the plane. New contrast appears as the depth increases, with the 1.67 nm focal depth showing weak moiré contrast and the 4.86 nm focal depth (≈5 nm of top STO thickness) clearly resolving the moiré periodic structure of the twisted interface (**Extended Data Fig. 4**). This establishes that we can selectively probe the interface region separately from the top and bottom membranes.

Importantly, this imaging allows us to distinguish Sr-CSL, Ti-CSL, and O-CSL sites (**Fig. 3b-d**, **Extended Data Fig. 4d**) within the large interface moiré cell. These commensurate sites are located at the interface plane where identical ions of the STO top and bottom layers are aligned vertically across the interface and exhibit moiré periodicity. In addition, we define a non-CSL region where there is no alignment of identical ions across the interface. **Fig. 3b** shows that the Ti-centered moiré supercell of the 10.4° twisted STO bilayer is composed of four types of CSL (Sr-, Ti-, O-, and non-), each of them having different local atomic structure at the interface. There is no significant atomic reconstruction between the rigid twisted atomic structures and the actual interfacial structure, well confirmed by the dynamical image simulation with the aid of the theoretical structural model (**Extended Data Fig. 5**).

Our major finding comes from comparing the electronic structure of these different sites at the twisted membrane interface using EELS with high depth resolution. **Fig. 3f** shows EELS Ti-$L_{2,3}$ spectra obtained from the twisted interface at each CSL region. It is seen that the spectral features of Ti ions in cation-centered CSL (Sr, Ti) regions are STO bulk-like with well-separated $e_g$ and $t_{2g}$ peaks at the $L_{2,3}$ edges, whereas O-CSL and non-CSL regions show spectral deviations, indicating electronic reconstruction dependent on the local interface structure. The Ti-$L_{2,3}$ edge peak broadening with the negative peak shift and proportionally decreased $e_g$-$t_{2g}$ energy loss difference at the O-CSL and non-CSL regions indicate a reduced Ti oxidation state in these regions compared to that in the cation-centered CSL[39-41]. The reduced Ti valence state is also evident from our O-K edge data, showing weakened intensities of the transitions to the O-$2p$ states hybridized with Ti-$3d$ and Sr-$4d$ orbitals due to $d$-band filling (**Extended Data Fig. 6a, b**).

At the same time, we observe no electronic reconstruction at the corresponding CSL regions away from the interface. This is evident from **Fig. 3g** which shows depth-resolved Ti-$L_{2,3}$ edge spectra for the O-CSL region. While the top and bottom STO layers demonstrate well-separated $t_{2g}$ and $e_g$ peaks, the interface layer exhibits notable spectral changes driven by the reduced Ti coordination. Similar features are also seen in the O-K edge spectra for the O-CSL region (**Extended Data Fig. 6c, d**), as well as in the Ti-$L_{2,3}$ and O-K edge spectra for non-CSL region (**Extended Data Fig. 6e, f**), showing that the local electronic reconstructions are confined to the twisted interface.

All these results demonstrate that the reduced coordination of the Ti atoms at the interface due to missing nearest neighbor oxygen atoms in $TiO_6$ octahedra at the interfacial O-CSL and non-CSL regions lead to the reduced Ti oxidation state compared to that at the interfacial Ti- and Sr-CSL regions. This modulation of the Ti oxidation state within the moiré cell is seen from **Fig. 3e** showing the measured 2D map of the energy difference between the $e_g$ and $t_{2g}$ peaks in the Ti-$L_3$ edge spectra. The mapping clearly indicates that the $e_g$ - $t_{2g}$ energy difference is at its nominal value near the cation-centered CSL, while it is reduced near the O-CSL and non-CSL regions. These EELS mapping results (**Fig. 3e**) are in good agreement with the Bader charge mapping (**Fig. 1f**), indicating that the electronic behavior at the twisted STO bilayer interface is controlled by the local atomic registry in the moiré lattice.

We note that oxygen vacancies can be ruled out as the source of the observed electronic reconstructions at the twisted interface. This follows from the elemental quantification by EELS at the interface, showing that all CSL regions remains stoichiometric to $TiO_2$ (**Extended Data Fig. 7**). In addition, our oxygen vacancy formation energy calculations reveal positive formation energies, demonstrating that spontaneous occurrence of oxygen vacancies is energetically unfavorable for all CSLs (**Extended Data Fig. 8**).

**Theoretical modelling**

The discovered correlation between the electronic state of Ti atoms and their atomic coordination at the twisted oxide interface that leads to the Ti valence state modulation within the moiré cell is accompanied by the formation of a flat band at the top of the valence band of the twisted STO bilayer. This prediction follows from our DFT calculations of the electronic structure of a 10.4° twisted STO bilayer that incorporates multiple variables reflecting our experiments (Methods). **Fig. 4a** shows the calculated band structure indicating the formation of flat band at a top valence band with an extremely narrow bandwidth. This flat band reflects a 2D localized state at the twisted STO interface exhibiting a localized charge density at the O- and non-CSL regions of the moiré cell, while no charge density at the Sr-and Ti-CSL regions (**Fig. 4b**). The cross-sectional view along the color bars in **Fig. 4b** indicates that the electron density is localized at the interfacial $TiO_2$ monolayer around the O- and non-CSLs (**Fig. 4c**) but absent at the Ti- and Sr-CSLs (**Fig. 4d**). It is noteworthy that the populated O-2$p$ orbitals, reflecting the charge density in **Fig. 4b**, are ordered around the CSL point, forming a zig-zag orbital-order pattern known for some manganites and cuprates[29,42]. This indicates that the local atomic configurations within the moiré supercell not only control the charge disproportionation but also produce orbital ordering.

The emergence of the flat band is driven by the same mechanism as the electronic charge modulation at the twisted STO interface. The reduced coordination at the O- and non-CSLs (**Fig. 1d, e**) forms a dangling bond at the Ti site which, on one hand, leads to a reduced Ti valence and, on the other hand, produces a localized electronic state. This results in a clear correlation between the charge density associated with the flat band (**Fig. 4a**) and the Bader charge reflecting the Ti oxidation state (**Fig. 1f**).

We, furthermore, predict that the varying local atomic registry drives the modulation of electronic potential within the moiré supercell (**Extended Data Fig. 9**). While an alternating potential with moiré periodicity in twisted 2D van der Waals materials is known to produce interesting phenomena, such as, e.g., the emergence of flat bands in magic-angle twisted bilayer graphene[43], twisted oxide heterostructures exhibit an additional complexity arising from potential variation within the moiré supercell itself. This new degree of freedom may induce novel electronic phases and phenomena which have yet to be explored.

**Conclusions and outlook**

Our work uncovers a new degree of freedom in complex-oxide moiré heterostructures which does not exist in twisted van der Waals systems. Due to multiatomic unit cell and strong bonding across the twisted interface, moiré complex oxides exhibit an interfacial electronic charge state that varies within the moiré supercell following the local atomic registry. This is revealed by the interfacial 2D moiré charge modulation, which we discover through direct atomic imaging and electronic spectroscopy using depth-resolved STEM in an STO twisted membrane bilayer designed to generate a stable moiré supercell with internal local atomic alignments. Our theoretical modelling demonstrates a clear correlation between local atomic coordination and the oxidation state of Ti atoms at the twisted interface which we explain within the CSL framework. The local atomic structure differences at the moiré interface drives selective electron occupation and quantum confinement.

The observed moiré phenomena driven by local atomic registry at the twisted oxide interface may have broader implications. For example, the predicted electronic flat band is caused by the reduced local atomic coordination within the moiré supercell, rather than an alternating moiré potential, as in van der Waals systems. As a result, the flat band does not require a small "magic" angle and can be probed with appropriate hole doping. The localized nature of this band signifies a possibility of strong electronic correlations and new electronic phases associated with it. Moreover, we envision new spin-dependent phenomena in moiré oxide materials. In the case of twisted STO, an unconventional $d^0$ magnetism driven by the spin splitting of the O-*p* bands may occur in response to hole doping, resulting in moiré-periodic spin density. In the case of oxides which are magnetic in the bulk form, the modulation of exchange and Dzyaloshinskii-Morya interactions at the interface may lead to new spin textures and complex magnetic states. The emerging vortex-antivortex patterns at the twisted oxide interfaces can further impact their electronic and magnetic properties, e.g., producing non-trivial multiferroic properties, which are yet to be studied.

Overall, our results demonstrate that the strong interactions across the twisted oxide interface, in conjunction with the formation of commensurate lattices at certain twist angles, lead to novel moiré phenomena driven by the interplay between the electronic and lattice degrees of freedom. This opens promising avenues to control the local electronic structure by the twist angle and paves new paths for interface engineering by oxide twistronics.

## Data Availability

The data supporting this study's findings are available from the corresponding authors on reasonable request

## Acknowledgements

S.Y.-C. acknowledges supports for this research by Korea Basic Science Institute (National Research Facilities and Equipment Center) grant funded by the Ministry of Education (2020R1A6C101A202) and this work was supported by the National Research Foundation of Korea (NRF) grant funded by the Korea government (MSIT) (RS-2024-00355591). C.B.E acknowledges support for this research through a Vannevar Bush Faculty Fellowship (ONR N00014-20-1-2844), and the Gordon and Betty Moore Foundations EPiQS Initiative, Grant GBMF9065. Fabrication of twisted membranes at the University of Wisconsin–Madison was supported by the US Department of Energy (DOE), Office of Science, Office of Basic Energy Sciences (BES), under award number DE-FG02-06ER46327. E.Y.T acknowledges support from the National Science Foundation through EPSCoR RII Track-1 program (NSF Award OIA-2044049). R.I. acknowledges support from JST FOREST (JPMJFR2033) and JSPS KAKENHI (JP24H00373). R.I. and N.S. acknowledge support from JST ERATO (JPMJER2202).

## Author Contributions

M.-S.K., K.L., C.B.E., and S.-Y.C. conceived the idea and designed the project. M.-S.K. carried out (S)TEM experiments and analyzed the microscopic and spectroscopic data under the supervision of S.-Y.C. C.B.E. K.L and K.T.E. fabricated the epitaxial thin films and freestanding membrane. R.I. performed depth-sectioning and EELS experiments and K.S.

conducted EELS experiment. T.M., E.Y.T., and N.A.S performed DFT calculations for electronic structure. M.S.R. measured transport and electrical properties. M.-S.K. and K.L. wrote the paper and all authors commented on it. S.-Y.C. and C.B.E. directed the overall project.

**Competing Interests Statement**

The authors declare no competing interests

# Figures

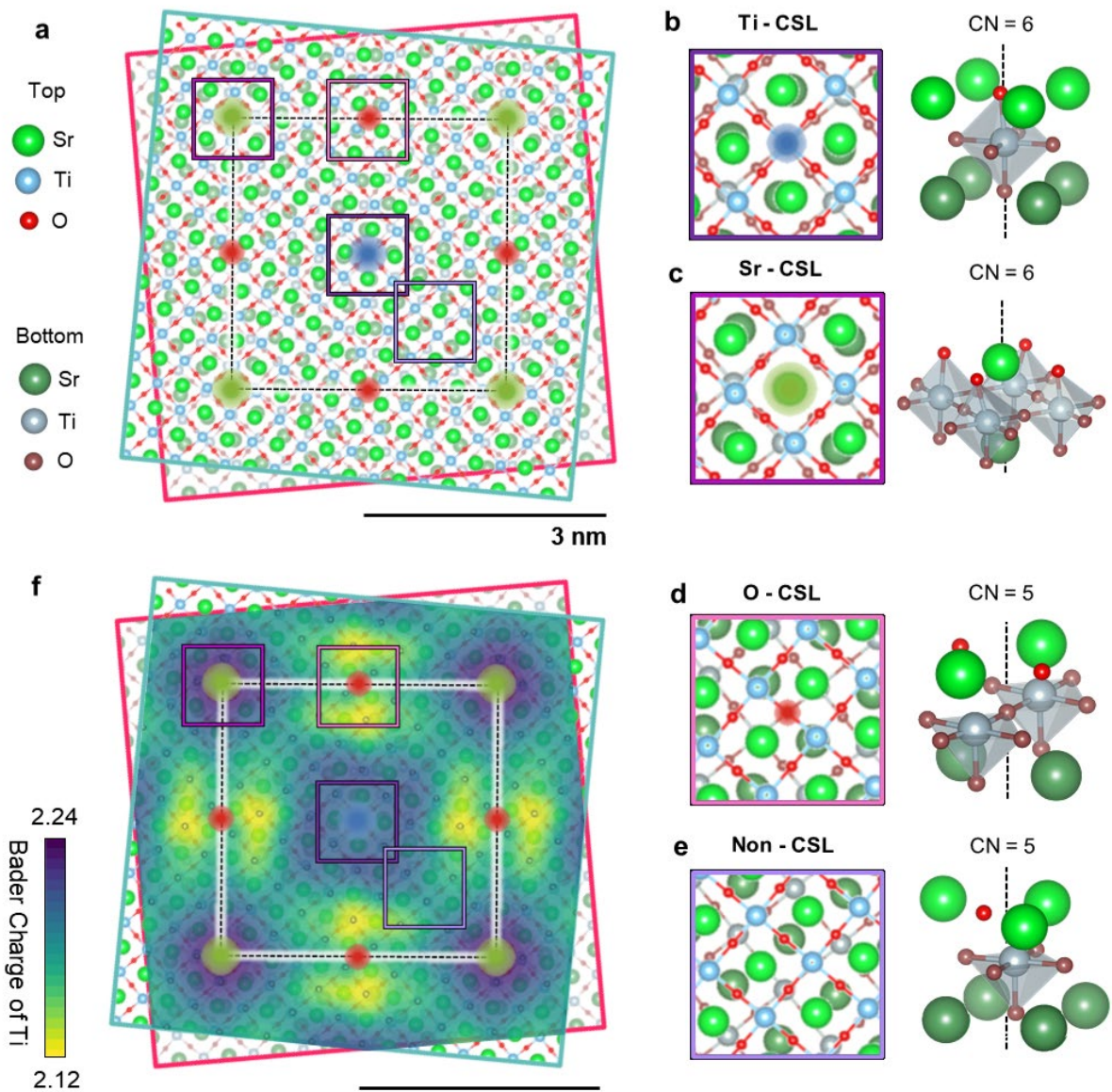

**Fig. 1 | Self-mediated moiré electronic lattice. a**, Schematic representation of commensurate moiré lattice of 10.4° twisted STO bilayers under $\Sigma$ 61 CSL boundary. The bright atomic structure with mint square and dark structure with pink square represent the top and bottom STO, respectively. The white dashed box represents a moiré supercell and CSL components are highlighted as violet (Sr), dark violet (Ti), red (O), pink (Non) color squares in **a**, **f**. **b-e**, Schematic of plan-view and interfacial atomic geometry for Sr-, Ti-, O-, and non-CSL region, respectively. The interfacial Ti-O bonding coordinates are dependent on the CSL and local lattice difference, inducing the charge disproportionation between 6- (Sr- and Ti-CSL, CN = 6) and 5-fold symmetric region (O- and Non-CSL, CN = 5). **f**, Bader charge calculation corresponding to the same region of a based-on charge disproportionation in moiré supercell.

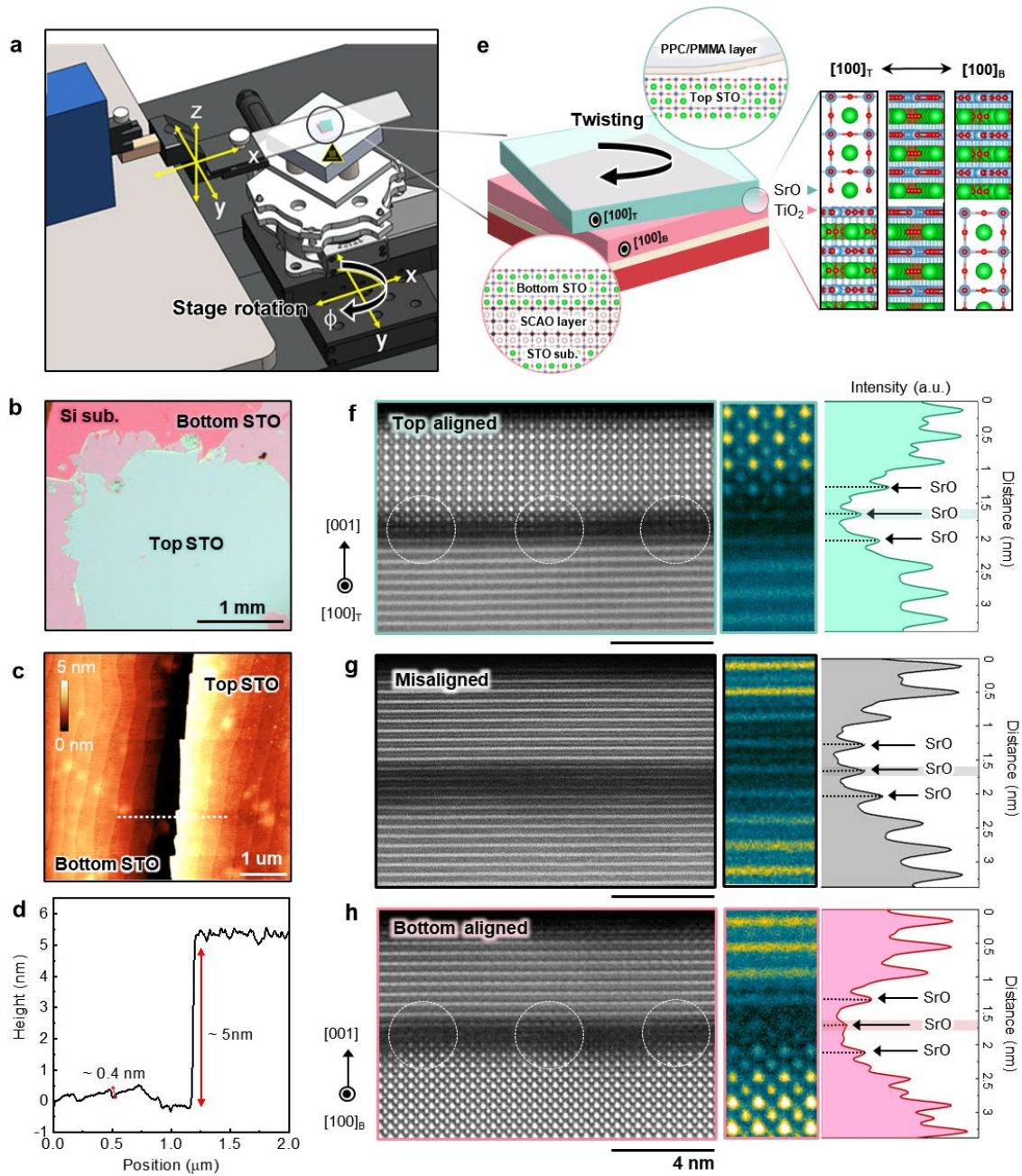

**Fig. 2 | Formation of twisted interfaces. a**, Schematic of twisting system for fabrication of twisted STO bilayers. **b**, Optical micrograph of twisted STO bilayers. **c**, Topographic image for boundary of top and bottom STO membrane investigated by AFM. **d**, Height profile from bottom to top STO membrane. **e**, Systematic representation of twisted STO bilayers and its interfacial structure by twist angles. Atomic-scale cross-sectional HAADF-STEM image (left), magnified image at bonded interface (middle), and line profile of the middle image (right) for **f**, top aligned, **g**, interface aligned, and **h**, bottom aligned zone axis, respectively. The difference in the tilt of the zone axis between the top and bottom alignment is 10.4°, equivalent to the twist angle, and the tilt angle of the interface zone axis has an intermediate value of about 5°.

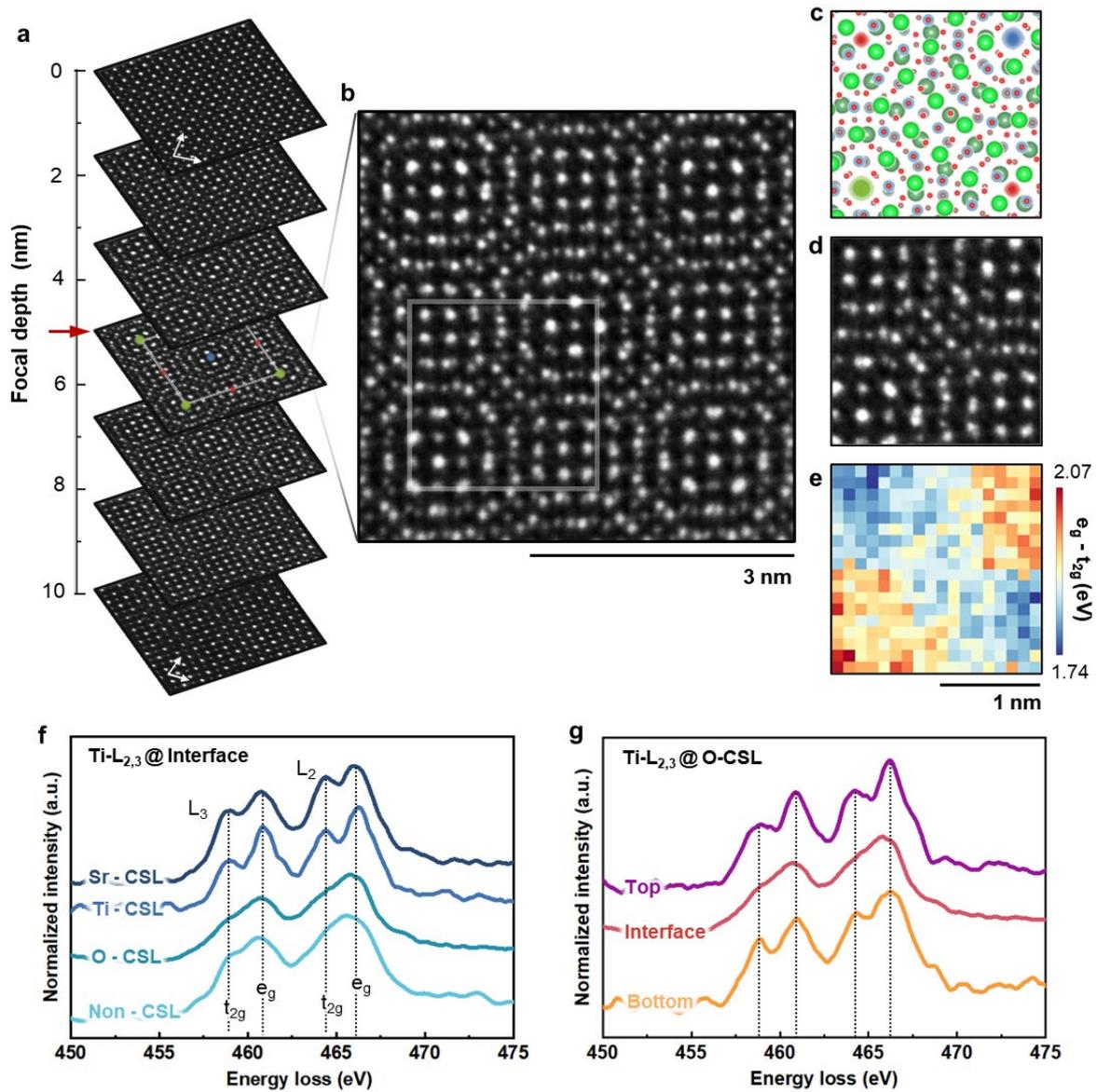

**Fig. 3 | Depth-resolved signatures of electron disproportionation. a**, A focal series of HAADF-STEM images from top to bottom region of 10.4° twisted STO bilayers. The white arrows in the image at top and bottom layer represent the in-plane direction of the STO. The components of moiré supercell are highlighted as green (Sr-), blue (Ti-) and red (O-) overlapped circles in **a**. The red arrow indicates the twisted interface. **b**, Atomic-scale HAADF image of moiré supercell for 10.4° twisted STO bilayers. **c-e**, Atomic structure, cropped HAADF-STEM image, and spectrum mapping of $e_g$-$t_{2g}$ difference in Ti-$L_3$ edge for a highlighted region in **b**, respectively. The significant reduction of $e_g$-$t_{2g}$ energy difference appears at O-CSL region, and slightly low $e_g$-$t_{2g}$ value at non-CSL whereas the Sr- and Ti-CSL retain its bulk-like Ti oxidation state (+4). **f**, EEL spectra for Ti-L at twisted interface obtained from each CSL region. **g**, Depth-resolved EEL spectra for Ti-L at O-CSL region.

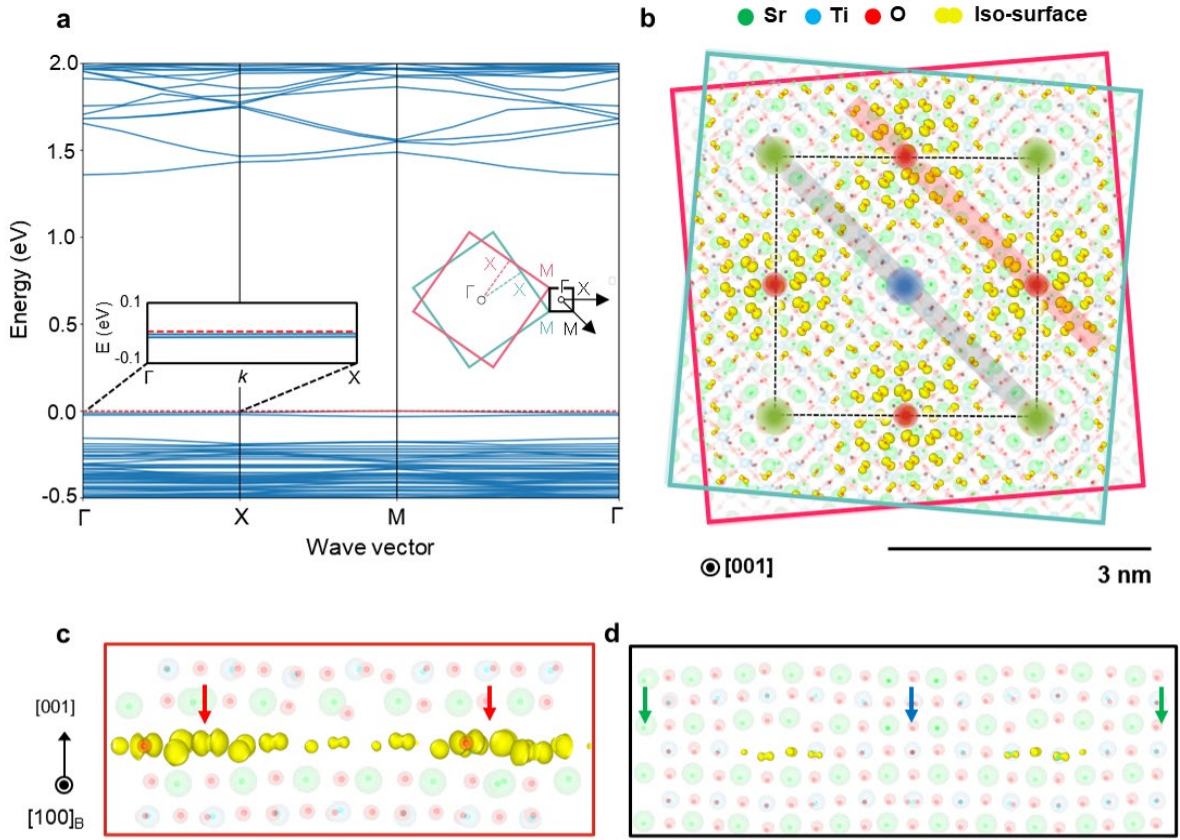

**Fig. 4 | Results of electronic structure calculations. a**, Band structure of a 10.4° twisted STO bilayer. The flat band appears along the Fermi level indicated by the red dashed line. Inset shows the 2D Brillouin zone of the twisted bilayer formed from the Brillouin zones of the two STO layers, with high symmetry *k*-points indicated. **b**, Top view of the distribution of the flat-band charge density of a 10.4° twisted STO bilayer represented by yellow-colored iso-surface. The localized charge density is located around the O- and non-CSL regions. **c**, **d** Side view of the distribution of the flat-band charge density along the red and black transparent lines in **a**, respectively. The solid red, green, and blue arrows represent O-, Sr-, and Ti-CSL, respectively.

22